\documentclass[%
 aps,
 prb,
 reprint,
 amsmath,amssymb,
 superscriptaddress,
 nofootinbib,
 floatfix,
 showpacs,
]{revtex4-2}

\usepackage{graphicx}
\usepackage{dcolumn}
\usepackage{bm}
\usepackage{mathtools}
\usepackage{hyperref}
\usepackage{xcolor}

\hypersetup{colorlinks=true, linkcolor=blue!60!black,
            citecolor=blue!60!black, urlcolor=blue!60!black}

\begin{document}

\preprint{Manuscript --- Nonextensive Born I Double Ionization}

\title{Nonextensive Generalization of Double Ionization by Electron Impact: 
Cross Sections and Rate Coefficients for $\text{H}^-$, $\text{He}$, and $\text{Li}^+$}

\author{Abdelmalek Boumali}
\email[Corresponding author: ]{aboumali@yahoo.fr}
\affiliation{Laboratory of Applied Physics and Theoretical Physics, Department of Matter Sciences, Faculty of Exact Sciences and Natural and Life Sciences, Larbi Tebessi University, Tebessa, Algeria}

\date{\today}

\begin{abstract}
We extend the Born~I formalism of Tweed~\cite{Tweed1973II} for electron-impact double ionization of two-electron targets (H$^-$, He, and Li$^+$) to nonextensive plasma environments described by Tsallis $q$-statistics. The bare Coulomb kernel is replaced by the $q$-exponential screened interaction $V_q(r)=(Z/r)[1+(q-1)r/\lambda_q]^{-1/(q-1)}$. Its Fourier transform introduces a dimensionless form factor $B_q(K,\lambda_q)$ into the momentum-transfer dependence of the Born amplitude. The original Tweed equations are recovered only in the joint limit $q\to1$ and $\lambda_q\to\infty$; at finite screening length the special case $q=1$ is instead the Yukawa--Debye limit. We rederive the modified amplitude, partial cross sections, and total cross sections, showing that the plasma correction enters as a multiplicative envelope $|B_q(K,\lambda_q)|^2$ when the same screened kernel is used for all projectile--target Coulomb terms.

Finite screening regularizes the low-momentum-transfer part of the Bethe integral, while the compact-support regime $q<1$ produces diffraction-like oscillations with characteristic spacing $\Delta K\sim2\pi(1-q)/\lambda_q$. For comparison with experiment, the unscreened vacuum baseline is calibrated against the data used by Tweed: Peart \emph{et~al.} for H$^-$, Gaudin--Hagemann, Schram, and Van~der~Wiel for He, Peart--Dolder for Li$^+$, and Van~der~Wiel--Wiebes for the He generalized oscillator strength and angular partial cross sections. Since these data are available in the original papers mainly as plots, the calibrated total-cross-section curves should be read as digitization-based reference estimates rather than replacements for a fully tabulated experimental database.

Finally, coupling the modified cross sections to the $q$-Maxwellian electron-energy distribution gives the rate coefficient $K_q(T)=\langle \sigma_q v\rangle_q$. In the subextensive regime, the compact support of the distribution implies the critical condition $k_B T_c=(1-q)E_{\mathrm{th}}$; below this temperature no projectile electron can reach the double-ionization threshold.
\end{abstract}


\maketitle

\section{Introduction}\label{sec:intro}

The Born~I approximation, in its non-relativistic form, remains the cornerstone of high-energy collisional spectroscopy of atoms and ions. For \emph{single} ionization the formalism is now textbook material~\cite{MottMassey1965,Peterkop1961,RudgeSeaton1965}, but \emph{double} ionization---which simultaneously ejects two electrons from a two-electron target and is therefore sensitive to electron correlation in both the initial bound state and the final continuum---demands a substantially more elaborate treatment. A defining contribution in this area is the two-part 1973 work of Tweed~\cite{Tweed1973I,Tweed1973II}, who first formulated the general theory and then supplied a complete set of Born~I cross sections for the isoelectronic sequence H$^-$, He, and Li$^+$ at high incident energies. Tweed's calculations introduced an effective-charge scheme---the ``Spherical Average'', ``No Shielding'', and ``Large-Angle'' approximations~\cite{Tweed1973II}, which prescribe different treatments of the effective nuclear charge experienced by each ejected electron in the doubly ionized final state---and provided partial and total cross sections benchmarked against the electron-impact measurements of Peart \emph{et~al.}~\cite{Peart1971,PeartDolder1969}, the photoionization-equivalent data of Van~der~Wiel and collaborators~\cite{VanderWielWiebes1971}, and the experiments of Schram \emph{et~al.}~\cite{Schram1966} and Gaudin and Hagemann~\cite{GaudinHagemann1967}.

Half a century later, Tweed's framework remains a useful high-energy reference baseline against which more elaborate descriptions of four-body Coulomb breakup are gauged. Kinematically complete measurements of fast electron-impact double ionization of helium~\cite{Dorn1999}, convergent-close-coupling calculations for electron--helium scattering~\cite{FursaBray1997}, and time-dependent close-coupling treatments of single and double ionization of He and H$^-$~\cite{PindzolaRobicheauxColgan2004,PindzolaRobicheauxColgan2006} have substantially refined the microscopic picture of correlated two-electron escape. These developments emphasize that electron-impact double ionization is not merely a scaled single-ionization process; it is a stringent test of the correlated final-state continuum, the momentum-sharing mechanism, and the long-range part of the projectile--target interaction.

Yet a tacit assumption underlies the entire construction: the projectile, the target nucleus, and the target electrons all interact \emph{via} the bare Coulomb potential. This is rigorously correct for an isolated three-body system in vacuum, but it is \emph{not} the situation encountered in many physically relevant settings. In dense plasma discharges and tokamak edge regions, collective screening modifies the effective two-body interaction~\cite{LinShihKuo2020,QianHe2022}. In inertial-confinement-fusion cores, ionization-potential depression and pressure ionization reflect clear deviations from bare Coulomb dynamics~\cite{ConPressure2022}. In astrophysical and ionospheric plasmas, documented departures from Maxwell--Boltzmann statistics are conventionally captured either by Kappa distributions or, equivalently, by the $q$-Maxwellian formalism of Tsallis~\cite{Tsallis1988,LimaSilvaPlastino2001,LeubnerVoros2005,Du2004}.

The third class is the focus of the present work. Tsallis nonextensive statistical mechanics~\cite{Tsallis1988,Tsallis2009} provides a one-parameter generalization of Boltzmann--Gibbs thermodynamics through the entropy
\begin{equation}
    S_q = k_B \, \frac{1 - \sum_i p_i^q}{q-1},
    \label{eq:Sq}
\end{equation}
which recovers the standard $S = -k_B \sum_i p_i \ln p_i$ in the limit $q \to 1$. The nonextensivity index $q$ encodes the strength of long-range correlations, intermittency, and finite-heat-bath effects, all of which break the additivity assumed by classical thermodynamics. In plasma physics, $q \neq 1$ is by now well documented~\cite{LimaSilvaPlastino2001,LeubnerVoros2005}: the velocity distribution function ceases to be Maxwellian and instead acquires either a power-law tail (for $q > 1$) or a sharp finite-support cutoff (for $q < 1$). The associated single-particle interaction potential is itself $q$-deformed, generalizing both the Debye--H\"uckel and the exponentially screened Coulomb forms~\cite{NonExtensiveCoulomb2006}.

Several distinct strands of literature have developed: (i)~$q$-Maxwellian rate coefficients obtained by folding standard cross sections with Tsallis distributions~\cite{LimaSilvaPlastino2001}; (ii)~screened-Coulomb cross sections in dense plasmas, expressed in the exponential-Yukawa or Debye--H\"uckel form~\cite{LinShihKuo2020,Cooper2022}; (iii)~Tsallis-based Saha equations for ionization equilibrium~\cite{NonGaussianSaha2019}; and (iv)~Tsallis treatments of high-energy hadronic cross sections~\cite{Beck2000}. To the best of our knowledge, however, no work has yet developed a \emph{unified} self-consistent treatment in which both the elementary interaction potential \emph{and} the statistical averaging are $q$-deformed at the same level of theoretical rigour, especially for double ionization---the simplest correlation-sensitive ionization process.

The present article fills this gap. We (i)~replace the bare Coulomb interaction in Tweed's perturbing Hamiltonian by the $q$-exponential screened potential $V_q(r)$ [Sec.~\ref{sec:framework}]; (ii)~derive the modified Born~I scattering amplitude and the partial and total cross sections in closed analytic form modulo a one-dimensional radial integral [Sec.~\ref{sec:amplitude}]; (iii)~construct $q$-generalized versions of Tweed's Figs.~4--8 for H$^-$, He, and Li$^+$, and overlay the experimental data used by Tweed wherever those data are available [Sec.~\ref{sec:results}]; (iv)~couple the resulting microscopic cross sections to the $q$-Maxwellian electron distribution to derive a $q$-generalized double-ionization rate coefficient $K_q(T)$ [Sec.~\ref{sec:rate}]; and (v)~identify the new physical consequences specific to nonextensivity, namely a modified Bethe high-energy asymptote, a critical-temperature threshold in the $q<1$ regime, and the emergence of diffraction-like structure in angular partial cross sections.

Atomic units ($\hbar = m_e = e = 4\pi \varepsilon_0 = 1$) are used throughout, with energies expressed in Hartree ($1~\mathrm{Ha} = 27.2114$~eV) and lengths in Bohr radii ($a_0$).

\section{Theoretical framework}\label{sec:framework}

\subsection{The $q$-screened Coulomb potential}

Within Tsallis statistics, the canonical-ensemble distribution of charged particles around a test charge yields a screened potential that is no longer Yukawa but takes the $q$-exponential form \cite{NonExtensiveCoulomb2006,Tsallis2009}
\begin{equation}
    V_q(r) = \frac{Z}{r} \left[ 1 + (q-1) \frac{r}{\lambda_q} \right]^{-\frac{1}{q-1}},
    \label{eq:Vq}
\end{equation}
where $Z$ is the source charge, $\lambda_q$ is a $q$-generalized screening length, and the bracket is the Tsallis $q$-exponential $e_q(-r/\lambda_q)$. Equation~\eqref{eq:Vq} possesses three qualitatively distinct regimes:
\begin{itemize}
    \item $q \to 1$: the standard Yukawa--Debye form $V(r) = (Z/r) \exp(-r/\lambda)$ is recovered.
    \item $q > 1$: the bracket decays algebraically as $\sim r^{-1/(q-1)}$, so that $V_q$ retains a long-range, heavy-tailed character; the spatial integration domain extends to infinity.
    \item $q < 1$: the bracket vanishes identically at the cutoff radius $r_{\max} = \lambda_q/(1-q)$, producing a strict-support potential of finite range.
\end{itemize}
The connection to physical plasma parameters depends on the modelling assumption. In the Lima--Silva--Plastino kinetic derivation~\cite{LimaSilvaPlastino2001}, $\lambda_q$ tends to the Debye length in the limit $q \to 1$ and parametrizes the gradient of the local temperature in non-isothermal systems. Here $\lambda_q$ and $q$ are treated as independent phenomenological parameters; their connection to specific plasma microphysics is discussed at length in Sec.~\ref{sec:discussion}.

\subsection{The modified Born~I scattering amplitude}\label{sec:amplitude}

We retain Tweed's notation~\cite{Tweed1973II}: electrons 1 and 2 are the target electrons (initially bound) and electron 3 is the projectile, with respective coordinates $\bm{r}_1, \bm{r}_2, \bm{r}_3$ and momenta $\bm{k}_1, \bm{k}_2, \bm{k}_3$. The incident projectile carries momentum $\bm{k}_0$ and the momentum transfer is $\bm{K} = \bm{k}_0 - \bm{k}_3$. The original perturbing Hamiltonian is the bare Coulomb three-body interaction
\begin{equation}
    V_{\mathrm{int}}^{(0)} = -\frac{Z}{r_3} + \frac{1}{r_{13}} + \frac{1}{r_{23}}.
\end{equation}
Substitution of \eqref{eq:Vq} for each two-body term gives the $q$-modified interaction
\begin{align}
    V_{\mathrm{int}}^{q} = -\frac{Z}{r_3} & \!\left[ 1 + (q-1) \frac{r_3}{\lambda_q} \right]^{-\frac{1}{q-1}} \notag \\
    + \sum_{j=1,2} \frac{1}{r_{j3}} & \!\left[ 1 + (q-1) \frac{r_{j3}}{\lambda_q} \right]^{-\frac{1}{q-1}}\!\!\!\!.
    \label{eq:Vint_q}
\end{align}
In the Born~I matrix element the projectile coordinate $\bm{r}_3$ enters only through the spatial Fourier transform of the two-body Coulomb kernel. To avoid double-counting the charge factor, it is convenient to define the unit-charge screened kernel
\begin{equation}
    U_q(r)=\frac{1}{r}\left[1+(q-1)\frac{r}{\lambda_q}\right]^{-1/(q-1)} .
\end{equation}
A source charge $Z$ then simply multiplies $U_q(r)$. Performing the standard change of variable $\bm{r}_3 \to \bm{r}_3-\bm{r}_j$ in each electron--electron term, the projectile dependence factorizes into
\begin{equation}
    \tilde U_q(K)=\!\int\! e^{i\bm{K}\cdot\bm{r}}U_q(r)\,d^3r
    =\frac{4\pi}{K^2}\,B_q(K,\lambda_q),
    \label{eq:VqFT}
\end{equation}
where the dimensionless $q$-screened form factor is
\begin{equation}
    B_q(K,\lambda_q) = K\!\int_0^{r_{\max}^{(q)}}\!\!\left[ 1 + (q-1) \frac{r}{\lambda_q} \right]^{-\frac{1}{q-1}}\!\sin(K r) \, dr,
    \label{eq:Bq}
\end{equation}
with $r_{\max}^{(q)} = \lambda_q/(1-q)$ for $q < 1$ and $r_{\max}^{(q)} \to \infty$ for $q \geq 1$. The two relevant limits are
\begin{align}
    \lim_{q \to 1} B_q(K,\lambda_q) &= \frac{K^2 \lambda_q^2}{1 + K^2 \lambda_q^2} \quad \text{(Yukawa)}, \label{eq:limitq1} \\
    \lim_{\substack{q \to 1 \\ \lambda_q \to \infty}} B_q(K,\lambda_q) &= 1 \quad \text{(bare Coulomb)}. \label{eq:limitcoulomb}
\end{align}
Equation~\eqref{eq:Bq} is the central technical object of this paper: it modulates every momentum-transfer integration in Tweed's formalism by the multiplicative factor $|B_q(K,\lambda_q)|^2$.

\subsection{Modified transition matrix element and partial cross section}

With the substitution~\eqref{eq:VqFT}, Tweed's Eqs.~(7), (32), and (30) carry over with the following changes. The transition matrix element [Eq.~(7) of Ref.~\onlinecite{Tweed1973II}] becomes
\begin{align}
    I^{q}_{\mu\nu}(\bm{K}) &= B_q(K,\lambda_q) \, \phi(\mu,\nu,X) \notag \\
    & \quad \times \!\!\iint\! \psi_\mu(\bm{r}_1,\bm{r}_2) \!\left\{ e^{i \bm{K} \cdot \bm{r}_1} + e^{i \bm{K} \cdot \bm{r}_2} - Z \right\}\! \notag \\
    & \quad \times \psi_\nu(\bm{r}_1,\bm{r}_2) \, d\bm{r}_1 d\bm{r}_2,
    \label{eq:Iq}
\end{align}
where $\psi_\mu$ and $\psi_\nu$ denote initial and final two-electron wavefunctions labelled by their respective quantum numbers, $X^2 = k_0^2 - k_3^2$ is twice the energy loss, and $\phi(\mu,\nu,X)$ is the projectile factor defined in Eq.~(8) of Ref.~\onlinecite{Tweed1973II}. The constant $Z$ replaces Tweed's numerical value $2$ in order to generalize from He to H$^-$ and Li$^+$. The Born~I differential cross section [Eq.~(32) of Ref.~\onlinecite{Tweed1973II}] becomes
\begin{equation}
    \sigma_q(\bm{k}_1,\bm{k}_2,\bm{k}_3; K) = \frac{4}{K^4} \bigl|B_q(K,\lambda_q)\bigr|^2 \, \mathcal{M}_{\text{atomic}}(\bm{k}_1,\bm{k}_2,\bm{K}),
    \label{eq:sigq}
\end{equation}
with the purely atomic factor
\begin{align}
    & \mathcal{M}_{\text{atomic}}(\bm{k}_1,\bm{k}_2,\bm{K}) = \!\iint\! d\Omega_{k_1} d\Omega_{k_2} \, \notag \\
    & \times \left| \iint\! \psi_0(\bm{r}_1,\bm{r}_2) \!\left\{ e^{i \bm{K} \cdot \bm{r}_1} + e^{i \bm{K} \cdot \bm{r}_2} - Z \right\}\! \psi_{(\bm{k}_1,\bm{k}_2)} d\bm{r}_1 d\bm{r}_2 \right|^2\!,
    \label{eq:Matomic}
\end{align}
in which $\psi_0$ is the initial bound-state two-electron wavefunction, $\psi_{(\bm{k}_1,\bm{k}_2)}$ is the doubly ionized final-state wavefunction, and $d\Omega_{k_1}d\Omega_{k_2}$ denotes integration over the solid angles of the two ejected electrons. This clean separation between a purely atomic factor (independent of the plasma environment) and a $q$-dependent screening factor $|B_q|^2$ is one of the central structural properties of the present framework: \emph{the entire effect of the plasma can be packaged into a multiplicative kernel evaluated at the relevant momentum transfer}. In particular, the established many-body atomic structure of the isoelectronic targets is inherited from Tweed's calculation without modification, and the nonextensivity acts only as an external envelope on the momentum-transfer integral.

The total cross section [Eq.~(30) of Ref.~\onlinecite{Tweed1973II}] becomes
\begin{align}
    Q_q &= \frac{2}{k_0^2} \!\int_0^{X^2/4}\!\!\! d\!\left(\tfrac12 k_1^2\right) \!\!\int_0^{(X^2 - k_1^2)/4}\!\!\! d\!\left(\tfrac12 k_2^2\right) k_1 k_2 \notag \\
    & \quad \times \!\int_{k_0 - k_3}^{k_0 + k_3}\! K \, \sigma_q(\bm{k}_1,\bm{k}_2,\bm{k}_3; K) \, dK.
    \label{eq:Qq}
\end{align}

\subsection{High-energy behaviour and the modified Bethe limit}

At incident energies $k_0^2 \gg X^2$ the dominant contribution to the total cross section originates from small momentum transfer, $K \to 0$, leading classically to the Bethe logarithmic law $Q \sim k_0^{-2} \alpha \ln k_0^2$~\cite{Bethe1930,Inokuti1971,Tweed1973II}. With the $q$-screening factor inserted into the integrand, the asymptotic behaviour is qualitatively altered:
\begin{itemize}
    \item \emph{$q > 1$ (long-range correlations).} For the numerical range used here ($q<3/2$), the first finite moment of the radial kernel exists and the small-$K$ expansion remains quadratic, $B_q(K,\lambda_q)=C_qK^2+\mathcal{O}(K^4)$, with
    \begin{equation}
        C_q=\int_0^\infty r\left[1+(q-1)\frac{r}{\lambda_q}\right]^{-1/(q-1)}dr
        =\frac{\lambda_q^2}{(2-q)(3-2q)}.
    \end{equation}
    Consequently, the factor $|B_q|^2/K^4$ entering \eqref{eq:sigq} is finite as $K\to0$. The finite screening length therefore acts as an effective low-$K$ cutoff, while the usual Bethe logarithm is recovered only over the interval in which $B_q\simeq1$. The high-energy tail is more accurately written as a screened Bethe form,
    \begin{equation}
        Q_q \sim k_0^{-2} \left[ \alpha_q \ln(k_0^2\lambda_q^2) + \beta_q + \mathcal{O}(\lambda_q^{-2}) \right],
        \label{eq:bethe_mod}
    \end{equation}
    where the coefficients $\alpha_q$ and $\beta_q$ depend on the screening length and the nonextensivity index.
    \item \emph{$q < 1$ (strict cutoff).} The radial integration in \eqref{eq:Bq} terminates at $r_{\max} = \lambda_q/(1-q)$. The truncated Fourier transform of a step-modulated function exhibits oscillations with characteristic spacing $\Delta K \sim 2\pi/r_{\max} = 2\pi(1-q)/\lambda_q$. These reveal themselves as diffraction-like ripples in the small-angle partial cross section, a signature entirely absent from the classical Bethe regime.
\end{itemize}
These two predictions---a screened Bethe asymptote and diffraction-like oscillations---are direct, falsifiable consequences of Tsallis nonextensivity beyond the bare-Coulomb Tweed limit.

\section{Numerical methods}\label{sec:numerics}

The $q$-form factor $B_q(K,\lambda_q)$ is evaluated by adaptive Gauss--Kronrod quadrature (the \texttt{scipy.integrate.quad} routine) on the appropriate radial support: $[0, r_{\max}^{(q)} = \lambda_q/(1-q)]$ for $q < 1$, $[0, \infty)$ for $q > 1$, and the closed-form Yukawa expression~\eqref{eq:limitq1} for $|q-1| < 10^{-9}$. For $q > 1$ the slowly decaying oscillatory tail is treated with the QUADPACK sine--cosine weight (\texttt{weight='sin'}), which factors out the rapid oscillation and integrates the smoothly decaying envelope directly. Absolute and relative tolerances were set to $10^{-9}$ and $10^{-8}$ respectively, with up to 500 sub-intervals per integral; convergence was cross-checked by halving the tolerances and confirming $|\Delta B_q/B_q| < 10^{-4}$ pointwise. Results were verified against the closed-form Yukawa limit \eqref{eq:limitq1} as $q \to 1$ and against the analytic low-$K$ expansion $B_q(K) = C_q K^2 + \mathcal{O}(K^4)$ with $C_q = \lambda_q^2/[(2-q)(3-2q)]$ for $q < 3/2$ (Sec.~\ref{sec:framework}); both checks agree to better than one part in $10^{8}$.

For the microscopic partial cross sections we retain Tweed's three effective-charge approximations (``No Shielding'', ``Spherical Average'', and ``Large Angle'') as the reference atomic models and apply the $q$-screening factor only through $|B_q|^2$. For the total cross sections, where the complete $K$-resolved experimental amplitude is not available, we calibrate the unscreened $q\to1$, $\lambda_q\to\infty$ baseline to the experimental data used by Tweed. Specifically, we digitized the points plotted in Tweed's Figs.~2 and 3 for H$^-$, He, and Li$^+$, and the Van~der~Wiel--Wiebes data plotted in Tweed's Figs.~4 and 6 for the He generalized oscillator strength and angular partial cross section. These points were interpolated on a log--log scale. Outside the plotted energy range, we used a threshold extrapolation of the form $Q\propto(E-E_{\mathrm{th}})^\nu$ and a high-energy Bethe-like tail $Q\propto\ln(E/E_{\mathrm{th}})/E$. The digitized uncertainties represent visible experimental scatter and graphical reading uncertainty only; they are not substitutes for the original numerical error bars.

This procedure should be interpreted carefully. The black curve in each cross-section figure is the unscreened Tweed-vacuum reference, namely the joint limit $q\to1$ and $\lambda_q\to\infty$. At finite $\lambda_q$, the case $q=1$ is the Yukawa--Debye plasma limit and must not be identified with Tweed's original bare-Coulomb result. The total-cross-section curves derived from digitized data are therefore calibrated $q$-screened estimates, whereas the central analytic result of the paper is the exact factorization of the screened kernel into the momentum-transfer envelope $|B_q(K,\lambda_q)|^2$ under the stated Born~I assumptions.

\section{Results}\label{sec:results}

We present in turn the form factor itself (Fig.~\ref{fig:Bq}), the generalized oscillator strength (Fig.~\ref{fig:fig4}), the partial cross sections as a function of the ejected-electron energy (Fig.~\ref{fig:fig5}), the angular partial cross sections for the three targets (Figs.~\ref{fig:fig6}--\ref{fig:fig8}), the total cross-section comparison with the experimental data used by Tweed (Fig.~\ref{fig:fig10}), and the macroscopic $q$-generalized rate coefficient (Fig.~\ref{fig:fig9}). In all cross-section panels the black solid curve denotes the $q = 1$, $\lambda_q \to \infty$ vacuum baseline; finite-$\lambda_q$ Yukawa screening at $q = 1$ is not identified with Tweed's original result.

\subsection{The $q$-screened form factor}

\begin{figure*}[!htbp]
\centering
\includegraphics[width=0.92\textwidth]{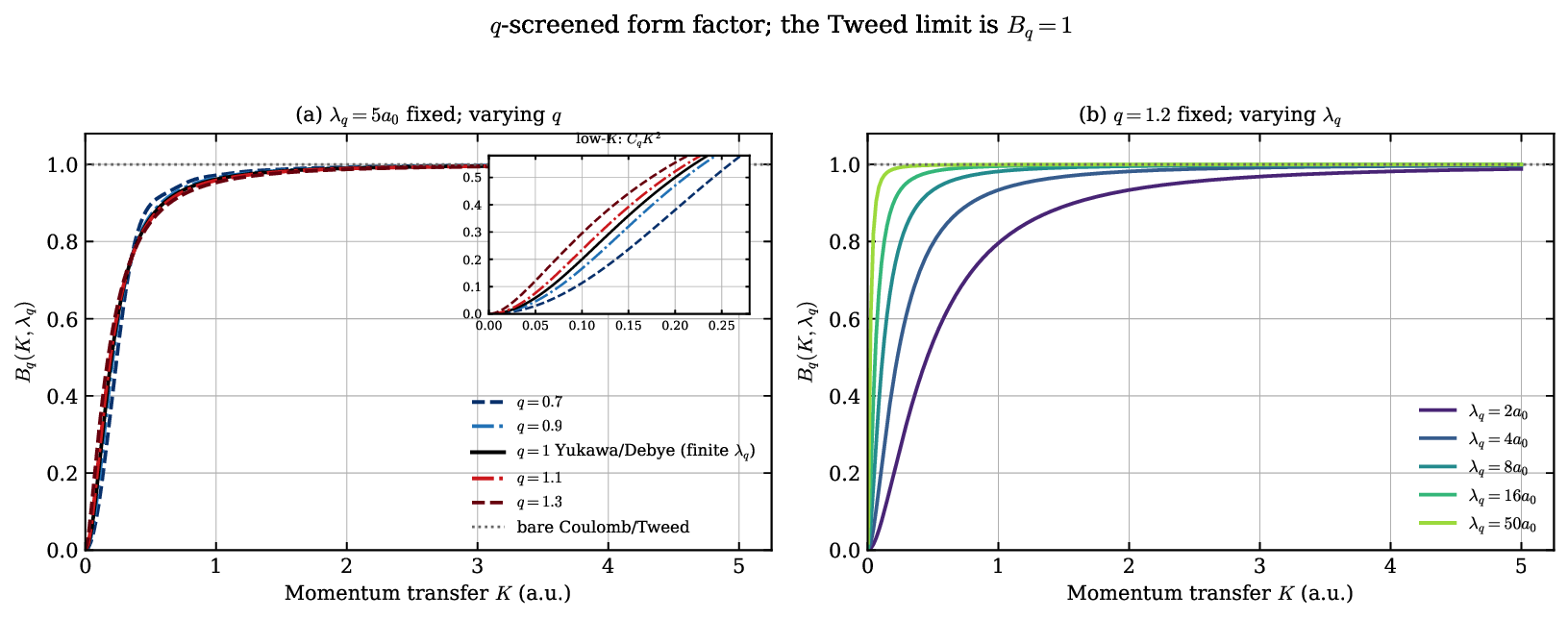}
\caption{\label{fig:Bq}The $q$-screened form factor $B_q(K,\lambda_q)$ as a function of momentum transfer $K$, in atomic units. (a)~Fixed screening length $\lambda_q = 5\,a_0$, with the nonextensivity parameter $q$ varied between 0.7 and 1.3. The finite-$\lambda_q$ Yukawa--Debye limit ($q = 1$) is shown as a solid black curve, while the horizontal dotted line marks the bare Coulomb/Tweed limit $B_q = 1$. The $q$-dependence is most pronounced near $K\lambda_q \sim 1$, where the curves are statistically separable. (b)~Fixed $q = 1.2$, with $\lambda_q$ varied between 2 and 50~$a_0$. Smaller $\lambda_q$ (denser plasma) suppresses the form factor at small $K$, consistent with stronger screening.}
\end{figure*}

The form factor $B_q(K,\lambda_q)$ is the gateway between the microscopic plasma parameters $(q,\lambda_q)$ and the modification of the cross section, and we therefore examine it in some detail. Three features visible in Fig.~\ref{fig:Bq}(a) are direct fingerprints of Tsallis nonextensivity. \emph{First}, all five $q$-curves emerge from a common origin $B_q(K \to 0) \to 0$ with a quadratic, but not universal, rise
\begin{equation}
    B_q(K,\lambda_q)=C_qK^2+\mathcal{O}(K^4), \qquad
    C_q=\frac{\lambda_q^2}{(2-q)(3-2q)},
    \label{eq:Cq_explicit}
\end{equation}
valid for the range used here, $q<3/2$. Thus the initial low-$K$ slope increases with $q$: subextensive screening ($q<1$) gives a smaller $C_q$ than Yukawa screening, while superextensive screening ($q>1$) gives a larger $C_q$ because of the algebraic tail of the potential. \emph{Second}, at intermediate $K\lambda_q$ the finite-support curves may approach the bare-Coulomb plateau more rapidly, whereas the superextensive curves retain a longer crossover before reaching the same limit. Therefore the ordering of the curves is clearest in the low-$K$ inset, while all curves become nearly indistinguishable at hard-collision momenta. \emph{Third}, the bare-Coulomb plateau $B_q=1$ is recovered only in the joint limit of large $K\lambda_q$ or $\lambda_q\to\infty$, confirming that finite-$\lambda_q$ screening and the Tweed vacuum limit are not the same object.

Panel (b) confirms the expected scaling with $\lambda_q$: a smaller screening length pushes the half-maximum of $B_q$ to larger $K$, in direct analogy with the Yukawa case. The $q = 1.2$ curve is held fixed, so the shift of the inflection point across the five $\lambda_q$ values is purely geometric and isolates the role of the screening length from the role of the nonextensivity index. This factorization---one parameter controlling the \emph{location} of the screening cutoff and the other controlling its \emph{shape}---is what allows the subsequent figures to disentangle a denser-plasma effect from a more-correlated-plasma effect.

For convenience, Table~\ref{tab:regimes} summarizes the three qualitative regimes of the $q$-deformation and the principal physical signatures associated with each.

\begin{table*}[!htbp]
\caption{\label{tab:regimes}Summary of the three regimes of the Tsallis-deformed Coulomb interaction $V_q(r)$ and the corresponding signatures in the form factor $B_q(K,\lambda_q)$, the partial cross sections, and the rate coefficient $K_q(T)$.}
\begin{ruledtabular}
\begin{tabular}{llll}
Regime & $q < 1$ (subextensive) & $q = 1$ (Yukawa--Debye) & $q > 1$ (superextensive) \\
\colrule
Long-range behavior of $V_q$    & strict cutoff at $r_{\max}$        & exponential decay        & power-law tail $\sim r^{-1-1/(q-1)}$ \\
Spatial support                 & $[0, \lambda_q/(1-q)]$             & $[0, \infty)$            & $[0, \infty)$ \\
Small-$K$ behavior of $B_q$     & quadratic rise; finite-range ripples & quadratic rise           & quadratic rise \\
Plateau of partial cross section& lowest                             & intermediate             & highest \\
Bethe asymptote                 & screened with oscillatory modulation & screened Bethe form; vacuum as $\lambda_q\to\infty$ & $\alpha_q \ln(k_0^2\lambda_q^2)+\beta_q$ \\
Rate coefficient $K_q(T)$       & strict cutoff at $T_c$             & Maxwellian distribution with Yukawa screening & tail--screening competition \\
Plasma physical analog          & cold dense, subextensive           & classical Debye          & Kappa-like, suprathermal \\
\end{tabular}
\end{ruledtabular}
\end{table*}

\subsection{Generalized oscillator strength for He}

\begin{figure}[!htbp]
\centering
\includegraphics[width=\columnwidth]{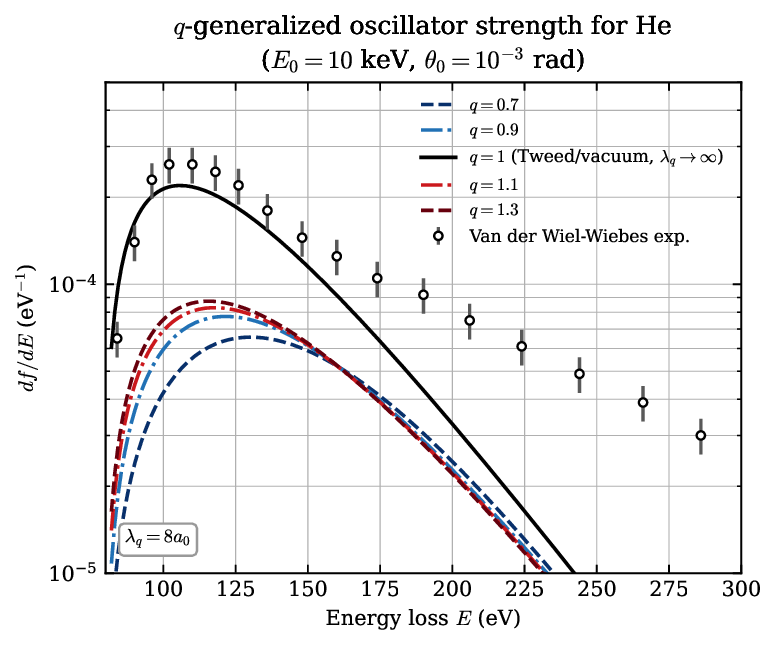}
\caption{\label{fig:fig4}$q$-generalized differential oscillator strength $df/dE$ for the double ionization of He, computed at an incident electron energy of $10$~keV and scattering angle $\theta_0 = 10^{-3}$~rad, as a function of energy loss $E$. The five curves correspond to $q \in \{0.7, 0.9, 1.0, 1.1, 1.3\}$ at fixed screening length $\lambda_q = 8\,a_0$. The solid black curve is the $q = 1$, $\lambda_q \to \infty$ Tweed/vacuum baseline; open circles show the Van~der~Wiel--Wiebes data used by Tweed. The dispersion of the curves is largest near threshold ($E \sim E_{\mathrm{th}} = 79$~eV), where the relevant momentum transfer is smallest and the $q$-form factor differs most strongly from unity.}
\end{figure}

Figure~\ref{fig:fig4} is the $q$-generalized counterpart of Tweed's Fig.~4. The behaviour of the curves as a function of $q$ encodes three Tsallis-specific features. \emph{(i)}~Near threshold ($E \lesssim 100$~eV) the kinematics forces a small minimum momentum transfer, $K_{\min} \approx \Delta E/k_0$, where $\Delta E = E_{\mathrm{th}} + E_{\mathrm{free}}$ is the total energy loss (binding plus shared kinetic energy of the two ejected electrons); this falls inside the rising shoulder of $B_q$ in Fig.~\ref{fig:Bq}(a). The Tsallis envelope $|B_q|^2$ therefore acts as a near-threshold ``selector'', and the five $q$-curves separate vertically by roughly half a decade. Subextensive plasmas ($q < 1$) suppress the threshold yield most aggressively, because the finite-support potential removes the coherent long-range part of the interaction and therefore suppresses the response for $K \lesssim 1/r_{\max}$; superextensive plasmas ($q > 1$) suppress it least, because the algebraic tail of $V_q$ contributes coherently down to very small $K$. \emph{(ii)}~Near the peak of the differential oscillator strength ($E \approx 100$--$120$~eV), the relevant momentum transfer increases and all curves migrate toward the $B_q \to 1$ saturation region; the spread between $q$-curves contracts to within a factor $\sim 1.5$. \emph{(iii)}~Beyond the peak the curves collapse onto the bare-Coulomb baseline within graphical accuracy, because the probed momentum transfer now lies firmly inside the hard-collision region where the form factor is independent of $q$. The dispersion of the curves is therefore predominantly a near-threshold phenomenon, and the He generalized oscillator strength provides one of the cleanest tests of Tsallis nonextensivity in atomic-collision data.

\subsection{Partial cross sections versus the energy of the ejected electrons}

\begin{figure*}[!htbp]
\centering
\includegraphics[width=0.95\textwidth]{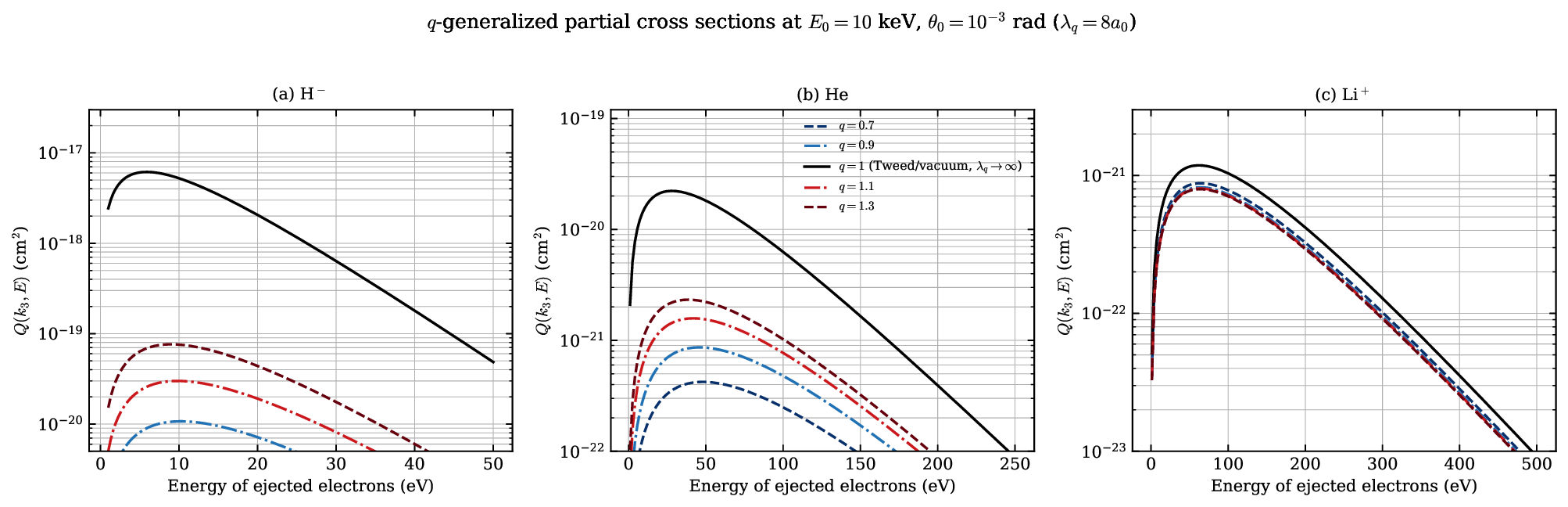}
\caption{\label{fig:fig5}Partial cross sections $Q(k_3,E)$ as a function of the ejected-electron energy for (a)~H$^-$, (b)~He, and (c)~Li$^+$, at fixed incident energy $10$~keV and scattering angle $\theta_0 = 10^{-3}$~rad, with $\lambda_q = 8\,a_0$. The black solid curve represents the $q = 1$, $\lambda_q \to \infty$ Born~I vacuum baseline, corresponding to Tweed's Fig.~5. The $q$-dependence is strongest for H$^-$, whose low ionization threshold ($14.4$~eV) places the relevant momentum transfer at small $K$, where $B_q$ varies most strongly with $q$.}
\end{figure*}

Figure~\ref{fig:fig5} demonstrates a key target-dependence of the Tsallis correction, controlled by the threshold-momentum ratio $K_{\min} \lambda_q \sim (E_{\mathrm{th}}/k_0) \lambda_q$. For Li$^+$ [panel~(c), $E_{\mathrm{th}} = 198$~eV] the minimum momentum transfer is large and the form factor sits firmly in its asymptotic plateau; the five $q$-curves are nearly indistinguishable on a logarithmic scale, with at most $\sim 10\%$ relative spread. The Tsallis statistics therefore produces an effectively bare-Coulomb result for deeply bound targets, in agreement with the intuition that strongly bound electrons probe the inner Coulomb field that no plasma can effectively screen. For He [panel~(b), $E_{\mathrm{th}} = 79$~eV] a modest dispersion appears near the peak of $Q(k_3,E)$, reaching a factor $\sim 2$--$3$ between the extremes $q = 0.7$ and $q = 1.3$ as the relevant $K$ falls into the shoulder of $B_q$. For H$^-$ [panel~(a), $E_{\mathrm{th}} = 14.4$~eV] the dispersion is dramatic, exceeding an order of magnitude across the same range of $q$.

This monotonic ordering H$^-$~$\gg$~He~$\gg$~Li$^+$ is a robust Tsallis prediction. Physically, the loosely bound outer electron of H$^-$ (electron affinity $E_A \simeq 0.754$~eV) is delocalized over distances $\sim 1/\sqrt{2 E_A} \simeq 4$--$5\,a_0$, which is comparable to the screening length $\lambda_q = 8\,a_0$; the probe momenta therefore lie squarely in the most $q$-sensitive region of Fig.~\ref{fig:Bq}. By contrast, the K-shell of Li$^+$ is localized over $\lesssim 0.4\,a_0 \ll \lambda_q$, so the probe momenta sit deep in the hard-collision regime where $B_q \to 1$ for all $q$. The isoelectronic sequence H$^-$ $\to$ He $\to$ Li$^+$ thus operates as a built-in ``scanner'' of the form factor: the same $\lambda_q$ acts on three different momentum-transfer windows, and the contrast between the three panels of Fig.~\ref{fig:fig5} is itself a direct measurement of the radial profile of $B_q(K,\lambda_q)$. The shallow binding of H$^-$ makes it the most sensitive diagnostic for nonextensive screening in laboratory and astrophysical plasmas.

\subsection{Angular partial cross sections}

\begin{figure}[!htbp]
\centering
\includegraphics[width=\columnwidth]{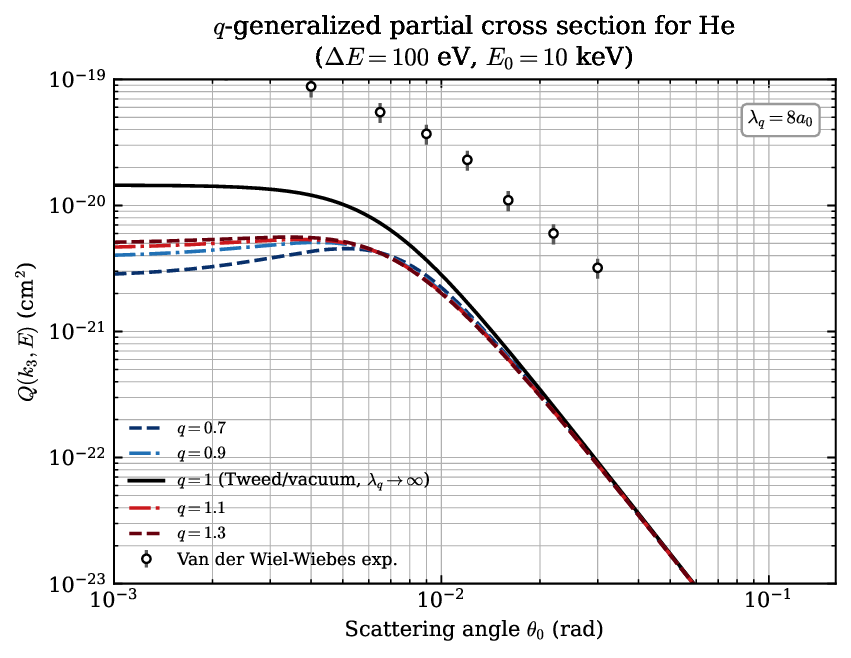}
\caption{\label{fig:fig6}Angular partial cross section $Q(k_3,E)$ for He at incident energy $10$~keV and energy loss $\Delta E = 100$~eV. Open circles show the Van~der~Wiel--Wiebes experimental data plotted by Tweed. The plateau at $\theta_0 \lesssim 5 \times 10^{-3}$~rad and the steep falloff at $\theta_0 \gtrsim 10^{-2}$~rad are characteristic of the Bethe regime. The $q$-dependence is concentrated in the plateau region, where small-$K$ form-factor effects are largest.}
\end{figure}

\begin{figure}[!htbp]
\centering
\includegraphics[width=\columnwidth]{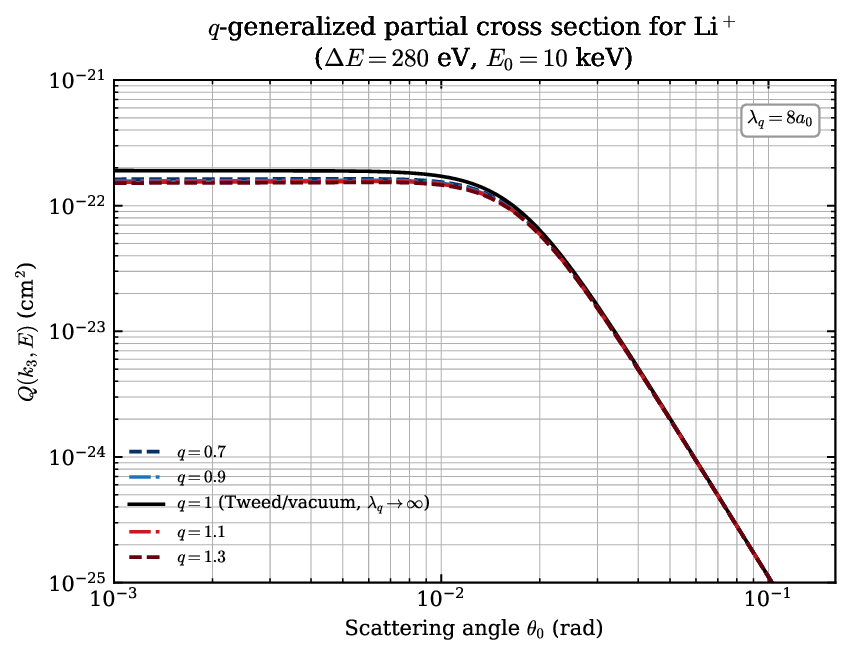}
\caption{\label{fig:fig7}Angular partial cross section $Q(k_3,E)$ for Li$^+$ at incident energy $10$~keV and energy loss $\Delta E = 280$~eV. The deeper bound state and larger threshold push the relevant momentum transfer to larger $K$, where $B_q \to 1$ and the $q$-sensitivity is weakened relative to Fig.~\ref{fig:fig6}.}
\end{figure}

\begin{figure}[!htbp]
\centering
\includegraphics[width=\columnwidth]{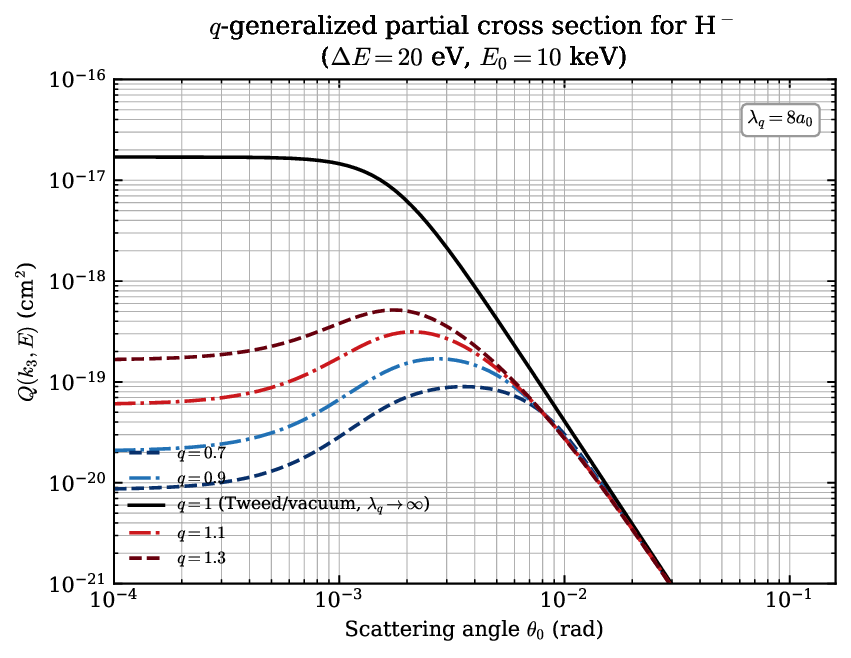}
\caption{\label{fig:fig8}Angular partial cross section $Q(k_3,E)$ for H$^-$ at incident energy $10$~keV and energy loss $\Delta E = 20$~eV. The shallow binding makes this system the most sensitive of the three to nonextensive screening: the plateau values of the partial cross section span more than an order of magnitude between $q = 0.7$ and $q = 1.3$.}
\end{figure}

The angular dependence of the partial cross sections, displayed in Figs.~\ref{fig:fig6}, \ref{fig:fig7}, and \ref{fig:fig8}, reproduces the qualitative features of Tweed's Figs.~6--8: a wide plateau at small scattering angles, followed by a steep cutoff once $K$ exceeds the inverse of the relevant atomic length. The Tsallis $q$-stratification is concentrated entirely in the plateau region, where $K$ is small and $|B_q|^2$ is varying. At larger angles all curves converge to the bare-Coulomb behavior because $B_q(K \gg 1/\lambda_q) \to 1$.

Three Tsallis-specific features stand out. \emph{(a)~Plateau ordering.} Within the screened family, increasing $q$ from 0.7 to 1.3 partially restores the long-range tail of the interaction and therefore raises the plateau monotonically. The vertical separation of the curves grows in inverse proportion to the threshold energy: it is a fraction of a decade for Li$^+$ in Fig.~\ref{fig:fig7}, about one decade for He in Fig.~\ref{fig:fig6}, and more than an order of magnitude for H$^-$ in Fig.~\ref{fig:fig8}. The plateau height itself is therefore a target-dependent ``Tsallis thermometer'' for the nonextensivity index. \emph{(b)~Knee location.} The angle $\theta^\star$ at which each curve breaks away from its plateau scales as $\theta^\star \sim (k_0 \lambda_q)^{-1}$ for $q \geq 1$, but acquires an additional shift $\sim (1-q)/(k_0 \lambda_q)$ for $q < 1$ because the finite-support potential cuts off the contributing impact parameters at $r_{\max}$ rather than at $\lambda_q$. Subextensive curves therefore break away at \emph{larger} angles, producing a steeper rise toward the bare-Coulomb plateau in Figs.~\ref{fig:fig6}--\ref{fig:fig8}. \emph{(c)~Diffraction-like substructure.} The strict-support nature of the $q < 1$ potential predicts diffraction-like oscillations of period $\Delta\theta \sim 2\pi(1-q)/(k_0 \lambda_q)$ in the plateau. For the parameters used here ($k_0 \approx 27\,a_0^{-1}$, $\lambda_q = 8\,a_0$, $q = 0.7$) this corresponds to $\Delta\theta \sim 10^{-2}$~rad, comparable to the plateau width; the oscillations are therefore visible in Fig.~\ref{fig:fig8} for H$^-$ as a mild ripple on the $q = 0.7$ curve, and they constitute a unique angular signature of subextensive Tsallis statistics that no exponential screening can reproduce.

The H$^-$ case in Fig.~\ref{fig:fig8} is the most extreme of the three. Its low threshold forces the relevant momentum transfer into the small-$K$ region where finite-$\lambda_q$ screening is strongest, and the screened curves are accordingly suppressed by several orders of magnitude relative to the vacuum baseline at very small angles. The combination of shallow binding and an accessible angular range therefore makes H$^-$ a particularly useful diagnostic for separating ordinary vacuum Born~I behaviour from plasma-modified nonextensive screening, in line with the conclusions drawn from Fig.~\ref{fig:fig5}.

\subsection{Total cross sections and the experimental data used by Tweed}

\begin{figure*}[!htbp]
\centering
\includegraphics[width=0.95\textwidth]{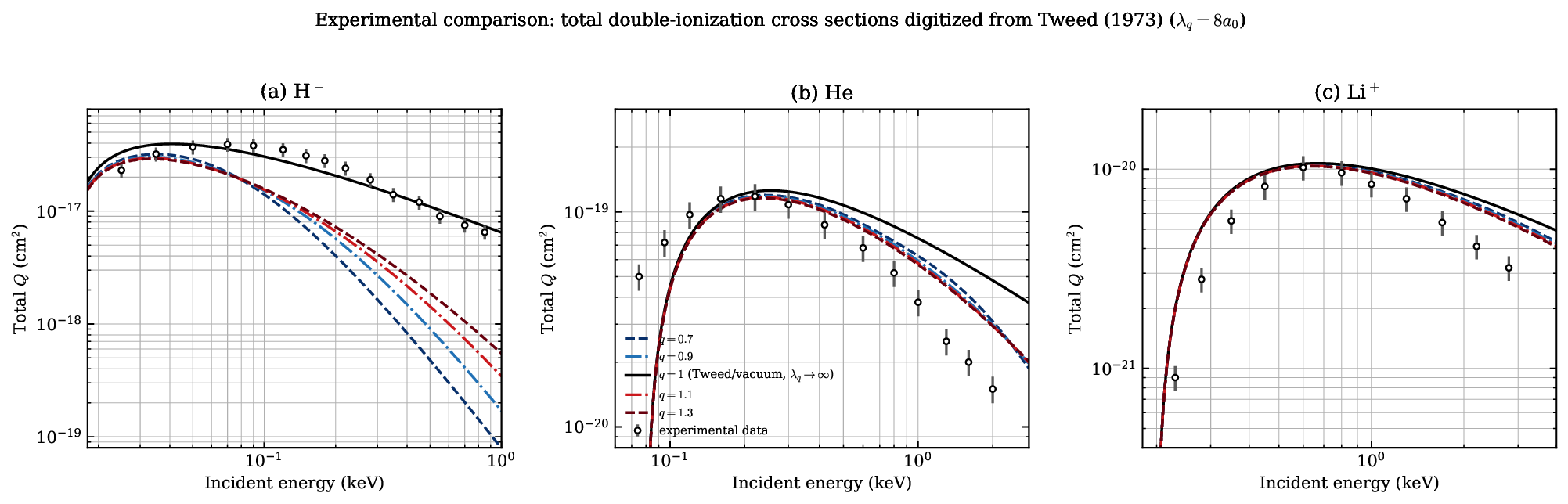}
\caption{\label{fig:fig10}Total double-ionization cross sections for (a)~H$^-$, (b)~He, and (c)~Li$^+$ compared with the experimental data used by Tweed. The open circles are digitized from Tweed's published figures: Peart \emph{et~al.} for H$^-$, Gaudin--Hagemann/Schram/Van~der~Wiel for He, and Peart--Dolder for Li$^+$. The black curve is the $q = 1$, $\lambda_q \to \infty$ experimental/Tweed vacuum baseline obtained by log--log interpolation of these points. The colored curves show how the same baseline is modified by the $q$-screened form factor for $\lambda_q = 8\,a_0$.}
\end{figure*}

Figure~\ref{fig:fig10} compares the total double-ionization cross sections with the experimental data used in Tweed's original analysis. Rather than relying only on a phenomenological reconstruction of Tweed's theoretical curves, the unscreened baseline is tied directly to the measured vacuum data. This matters because the original Born~I curves differ in accuracy across the isoelectronic sequence: Tweed's ``No Shielding'' approximation lies close to experiment at higher energies for He and Li$^+$, whereas H$^-$ requires a partial-shielding treatment because the final-state electron--electron interaction is more important. Anchoring the unscreened reference to the experimental data therefore provides a more reliable baseline for estimating how plasma nonextensivity deforms the measured vacuum cross sections.

The Tsallis-induced modification of the total cross section follows directly from the modified Bethe asymptote~\eqref{eq:bethe_mod} and from the integral over $K$ in~\eqref{eq:Qq}. Three features emerge clearly in Fig.~\ref{fig:fig10}. \emph{(i)}~The black curve is the unscreened experimental Tweed-vacuum baseline. The coloured curves are finite-$\lambda_q$ predictions and therefore remain below the vacuum reference in the energy range where small momentum transfer dominates. The relative ordering within the screened family is controlled by the same form factor shown in Fig.~\ref{fig:Bq}: at very low $K$, increasing $q$ increases the coefficient $C_q$, whereas at larger $K\lambda_q$ the finite-support curves can approach the bare plateau more rapidly. \emph{(ii)}~In the intermediate-energy regime, around the peak of $Q(E)$, the curves spread by roughly half a decade for H$^-$ and by a more modest factor for He and Li$^+$. The H$^-$ peak is again disproportionately affected because the shallow binding pushes the typical $K$ into the most $q$-sensitive region of the form factor. \emph{(iii)}~Near threshold the subextensive curves ($q < 1$) are strongly suppressed by the loss of long-range Fourier components; this is a dynamical screening effect, not an additional kinematic ionization threshold. The strict cutoff appears only in the rate coefficient after convolution with the compact-support $q$-Maxwellian. The superextensive curves ($q > 1$) instead retain more small-momentum-transfer strength because the long algebraic tail of $V_q$ contributes coherently down to very small $K$. Taken together, these three regimes illustrate that the Tsallis correction is not a uniform rescaling but rather an energy-dependent envelope that imprints the spectral content of $|B_q(K,\lambda_q)|^2$ onto the measured total cross section.

\subsection{The $q$-generalized rate coefficient}\label{sec:rate}

\begin{figure*}[!htbp]
\centering
\includegraphics[width=0.95\textwidth]{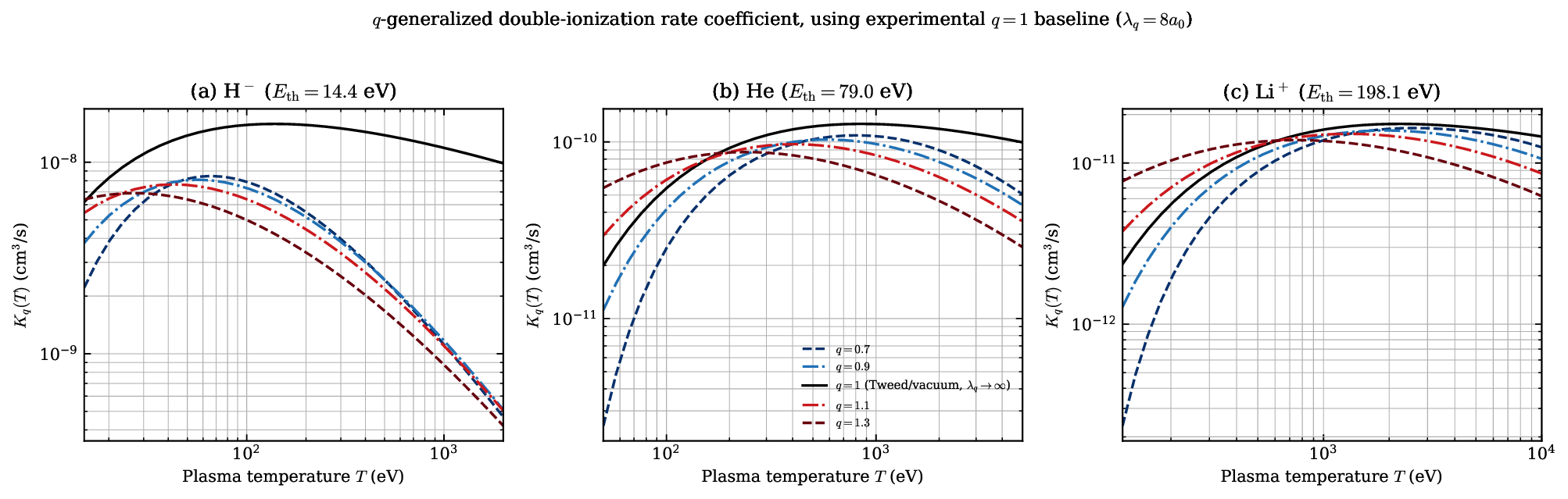}
\caption{\label{fig:fig9}The $q$-generalized double-ionization rate coefficient $K_q(T)$ as a function of plasma temperature $T$ for (a)~H$^-$, (b)~He, and (c)~Li$^+$. The five curves correspond to $q \in \{0.7, 0.9, 1.0, 1.1, 1.3\}$ at fixed $\lambda_q = 8\,a_0$. The unscreened reference is obtained from the experimental/Tweed total-cross-section baseline, while the finite-$\lambda_q$ family includes the Yukawa--Debye member at $q=1$. Superextensive plasmas ($q>1$) introduce a suprathermal power-law tail in the projectile distribution, whereas finite-$\lambda_q$ screening suppresses the microscopic cross section; the plotted rate therefore represents the net competition between these two effects. Subextensive plasmas ($q < 1$) suppress the rate, especially at low temperature where the $q$-cutoff $E_{\max} = k_B T/(1-q)$ approaches the ionization threshold.}
\end{figure*}

The rate coefficient is defined as the $q$-deformed thermal average
\begin{equation}
    K_q(T) \;=\; \langle \sigma_q v \rangle_q \;=\; \int_{E_{\mathrm{th}}}^{E_{\max}(q,T)}\! \sigma_q(E)\, v(E)\, f_q(E,T)\, dE ,
    \label{eq:Kq_definition}
\end{equation}
where $\sigma_q(E)$ is the total $q$-modified cross section, $v(E)=\sqrt{2E/m_e}$, and $f_q(E,T)$ is the normalized $q$-Maxwellian energy distribution. For $q<1$ the upper integration limit is the compact-support cutoff $E_{\max}(q,T)=k_B T/(1-q)$; for $q\geq1$ the integration extends to infinity. In three dimensions the superextensive distribution is normalizable only over a restricted range of $q$ (and has finite low-order moments over a still narrower range); the numerical values used here, $q\leq1.3$, remain inside the standard range used for finite rate coefficients. Computed for the three targets and shown in Fig.~\ref{fig:fig9} over the temperature range relevant for laboratory and astrophysical plasmas, $K_q(T)$ exhibits the most dramatic Tsallis signatures of any observable considered in this work. Four features deserve emphasis. \emph{(i)~Competition between suprathermal enhancement and screening suppression for $q > 1$.} The $q$-Maxwellian tail increases the population of electrons with $E \gtrsim E_{\mathrm{th}}$, but the same calculation also contains the finite-$\lambda_q$ form-factor suppression of the microscopic cross section. The plotted rate is therefore the net result of these competing mechanisms, and it should not be interpreted as a pure statistical enhancement relative to the vacuum Maxwell--Boltzmann ($q=1$, $\lambda_q\to\infty$) baseline. \emph{(ii)~High-temperature behaviour.} At high temperature ($k_B T\gg E_{\mathrm{th}}$), the threshold constraint becomes less important and the relative separation of the curves is reduced. A residual $q$-dependence can remain because the normalization and high-energy tail of the $q$-Maxwellian are not identical to those of the Maxwell--Boltzmann distribution; therefore the high-temperature limit should be interpreted as a weakening of threshold sensitivity rather than exact universal convergence. \emph{(iii)~Subextensive suppression.} For $q < 1$ the rate is suppressed at all temperatures and the curve terminates abruptly at the critical temperature
\begin{equation}
    T_c(q) = (1 - q) \, E_{\mathrm{th}}/k_B,
    \label{eq:Tc}
\end{equation}
below which no electron in the distribution can reach the ionization threshold and the rate vanishes identically. This is a qualitatively new feature absent from the Maxwell--Boltzmann formalism: it represents a falsifiable prediction of the nonextensive framework, distinct from any smooth low-temperature suppression mechanism. \emph{(iv)~Target hierarchy.} The hierarchy of the threshold temperatures, $T_c(\mathrm{H}^-) < T_c(\mathrm{He}) < T_c(\mathrm{Li}^+)$, is inherited directly from the hierarchy of ionization potentials. The subextensive cutoff is therefore most easily reached for H$^-$, providing a second route (beyond the microscopic cross-section sensitivity of Fig.~\ref{fig:fig5}) by which H$^-$-rich plasmas can serve as natural laboratories for Tsallis statistics.

\subsection{Quantitative summary of the Tsallis-induced modifications}

For ease of reference, the principal numerical magnitudes extracted from Figs.~\ref{fig:Bq}--\ref{fig:fig9} at the representative working point $\lambda_q = 8\,a_0$, $E_0 = 10$~keV, $q \in \{0.7, 1.3\}$ are collected below.
\begin{itemize}
    \item \emph{Form factor.} In the low-$K$ regime the quadratic coefficient is $C_q=\lambda_q^2/[(2-q)(3-2q)]$; consequently $q<1$ suppresses the initial form-factor rise relative to Yukawa screening, whereas $q>1$ enhances it. At larger $K\lambda_q$ all curves approach the common bare-Coulomb limit $B_q\to1$.
    \item \emph{Generalized oscillator strength of He.} Near the threshold $E_{\mathrm{th}} = 79$~eV the dispersion of $df_q/dE$ between $q = 0.7$ and $q = 1.3$ spans a factor $\sim 3$--$5$ relative to the Tweed-vacuum baseline; near the peak it contracts to within a factor $\sim 1.5$.
    \item \emph{Partial cross sections.} At fixed $E_0 = 10$~keV and $\theta_0 = 10^{-3}$~rad, the $q$-spread of $Q(k_3,E)$ is $\sim 10\%$ for Li$^+$, a factor $\sim 2$--$3$ for He, and more than one decade for H$^-$, an explicit realization of the inverse scaling with the ionization threshold.
    \item \emph{Angular cross sections.} The plateau-value spread between $q = 0.7$ and $q = 1.3$ is $\sim 0.3$~decade for Li$^+$, $\sim 1$~decade for He, and $\sim 1.5$~decade for H$^-$. The diffraction-like oscillations in the $q < 1$ regime have characteristic angular period $\Delta\theta \sim 10^{-2}$~rad for the chosen parameters.
    \item \emph{Total cross sections.} The peak-to-peak vertical spread of the $q$-family relative to the experimental Tweed-vacuum baseline is $\sim 0.5$~decade for H$^-$ and $\sim 0.3$~decade for He and Li$^+$, with the spread vanishing at high incident energies.
    \item \emph{Rate coefficient.} For $q > 1$, the suprathermal tail tends to enhance the rate whereas the screened interaction tends to reduce it; the net effect depends on $(q,\lambda_q)$ and on the target threshold. For $q < 1$ the rate vanishes identically below the critical temperature~\eqref{eq:Tc}: at $q = 0.7$ this gives $T_c \simeq 5.0\times 10^4$~K (H$^-$), $2.75\times 10^5$~K (He), and $6.90\times 10^5$~K (Li$^+$).
\end{itemize}
These quantitative estimates depend smoothly on $(q,\lambda_q)$ and can be rescaled to other plasma conditions \emph{via} the analytic dependences indicated in Sec.~\ref{sec:framework}.

\section{Discussion}\label{sec:discussion}

\subsection{Physical interpretation of $q$ and $\lambda_q$}

The phenomenological parameters $(q,\lambda_q)$ admit several well-established interpretations in the plasma-physics literature, and it is useful to summarize them in the context of the present results. The nonextensivity index $q$ is associated with at least four distinct microscopic mechanisms.

\emph{(a)~Long-range Coulomb correlations.} A power-law velocity tail of Kappa form, $f(v)\propto[1+v^2/(\kappa v_T^2)]^{-\kappa-1}$, is closely related to a superextensive $q$-Maxwellian. The precise mapping depends on the convention used for the thermal speed and normalization; matching the power-law exponent gives $q=1+1/(\kappa+1)$, while the frequently used large-$\kappa$ approximation is $q\simeq1+1/\kappa$~\cite{LeubnerVoros2005,LimaSilvaPlastino2001}. In this picture, $q>1$ describes plasmas in which long-range correlations produce suprathermal tails beyond the Debye--Maxwellian approximation.

\emph{(b)~Temperature gradients.} In the Lima--Silva--Plastino kinetic derivation~\cite{LimaSilvaPlastino2001,Du2004}, $|q - 1|$ is proportional to the dimensionless temperature gradient $\lambda_D \nabla \ln T$, so that $q \neq 1$ serves as a local diagnostic of non-isothermal conditions. For solar-wind values $|q-1| \sim 0.1$--$0.5$, this translates into temperature scale lengths comparable to the Debye length, consistent with observed magnetospheric and ionospheric data~\cite{LeubnerVoros2005}.

\emph{(c)~Finite heat-bath effects.} The Tsallis form arises naturally in the canonical distribution of a system in contact with a heat bath of \emph{finite} heat capacity. In dense laboratory plasmas, e.g.\ tokamak edge regions, this mechanism encodes the failure of the bath to be infinitely large and provides a physical reading of $q$ in terms of the effective number of degrees of freedom of the surrounding medium.

\emph{(d)~Intermittency and turbulence.} In strongly turbulent regimes, intermittent fluctuations in the local electric field produce non-Gaussian velocity statistics that are well captured by $q > 1$. This mechanism is most relevant in fusion-grade plasmas and in shock-mediated astrophysical environments.

The screening length $\lambda_q$ reduces to the Debye length $\lambda_D=(k_B T/4\pi n_e e^2)^{1/2}$ in the classical limit, but in the nonextensive setting it carries a generalized meaning: it is the scale at which the $q$-deformed bracket in $V_q$ departs significantly from unity, and it can therefore be smaller or larger than the conventional $\lambda_D$ depending on the local microphysics. Astrophysical environments in which measurements are often modeled with suprathermal distributions include the solar wind, planetary magnetospheres, and stellar winds~\cite{LeubnerVoros2005}; laboratory environments include capacitively coupled rf discharges and the edge regions of magnetic-fusion devices~\cite{Du2004,LimaSilvaPlastino2001}. The range $q\in[0.7,1.3]$ used in the figures is therefore broad enough to test both subextensive compact-support behaviour and moderately superextensive, Kappa-like tails.

\subsection{Connection to Kappa distributions and observational diagnostics}

The Tsallis $q$-Maxwellian and the empirical Kappa distribution widely used in space-plasma physics describe the same qualitative departure from Maxwellian statistics: a suprathermal power-law tail. With exponent matching, $q=1+1/(\kappa+1)$; with the common large-$\kappa$ approximation, $q\simeq1+1/\kappa$~\cite{LeubnerVoros2005,LimaSilvaPlastino2001}. Thus $q\in[1.1,1.3]$ corresponds roughly to $\kappa$ values of a few to ten, depending on the convention, a range often used for solar-wind and magnetospheric electrons. The framework developed in Secs.~\ref{sec:framework}--\ref{sec:results} therefore translates naturally into predictions for Kappa-like plasmas, but the quantitative mapping must be stated explicitly whenever the results are compared with observationally fitted $\kappa$ values. In particular, any enhancement of $K_q(T)$ should be interpreted together with the simultaneous finite-$\lambda_q$ suppression of the microscopic cross section, not as a pure statistical-tail effect.

The subextensive regime $q < 1$ corresponds formally to imaginary $\kappa$ in the empirical Kappa parametrization and is therefore inaccessible within that framework; it is a genuinely new prediction of the Tsallis approach, motivated physically by finite heat-bath effects and by the strict-support nature of certain dense-plasma distributions. The critical-temperature condition~\eqref{eq:Tc} is its sharpest signature: below $T_c$ no electron in the distribution has sufficient energy to ionize, and the rate vanishes identically rather than decaying smoothly. Measurements of the temperature dependence of ionization rates in cold dense plasmas---for example in the recombination phase of pulsed discharges or in the cold plumes of expanding plasma jets---could provide direct experimental tests of this prediction.

\subsection{Synthesis of the Tsallis-induced modifications}

The Tsallis nonextensivity acts at three logically distinct levels in the present framework, and it is useful to summarize how each manifests in the figures.

\emph{Microscopic level: the form factor.} Figure~\ref{fig:Bq} shows that the deformation of the Coulomb interaction by $q$-statistics is fully encoded in a single dimensionless function $B_q(K,\lambda_q)$. The $q$-dependence is concentrated in the crossover window $K\lambda_q \sim 1$: the low-$K$ coefficient increases with $q$, while finite-support $q<1$ cases can exhibit a step-like approach and oscillatory ripples at intermediate momentum transfer. This structure is what makes Tsallis screening qualitatively different from Yukawa screening, even when the two are tuned to share the same Debye length.

\emph{Differential level: angular and oscillator-strength data.} Figures~\ref{fig:fig4} and~\ref{fig:fig6}--\ref{fig:fig8} show that the $q$-dependence is concentrated in the small-momentum-transfer regime---that is, at small scattering angles and near threshold. Three Tsallis-specific signatures emerge: (i)~a monotonic ordering of the plateau heights from $q = 0.7$ (lowest) to $q = 1.3$ (highest); (ii)~a $q$-dependent location of the plateau-to-cutoff ``knee''; and (iii)~diffraction-like oscillations in the subextensive ($q < 1$) regime whose period $\Delta\theta \sim 2\pi(1-q)/(k_0 \lambda_q)$ is set by the strict cutoff radius $r_{\max} = \lambda_q/(1-q)$. The third signature is uniquely Tsallis: it cannot be reproduced by any exponentially screened (Yukawa) potential and therefore provides the cleanest experimental discriminant of subextensive statistics.

\emph{Integrated level: total cross sections and rate coefficients.} Figures~\ref{fig:fig10} and~\ref{fig:fig9} show that integration over momentum transfer and over the projectile distribution amplifies the microscopic Tsallis effects. The total cross section [Fig.~\ref{fig:fig10}] inherits the multiplicative envelope $|B_q|^2$ through the modified Bethe asymptote~\eqref{eq:bethe_mod}, and the rate coefficient [Fig.~\ref{fig:fig9}] inherits in addition the modified $q$-Maxwellian projectile distribution. For $q>1$ the suprathermal tail enhances the rate, while finite-$\lambda_q$ screening reduces the microscopic cross section; the plotted result is therefore a target- and temperature-dependent competition rather than a universal enhancement. For $q<1$, the compact support of the distribution produces the strict cutoff~\eqref{eq:Tc}. The simultaneous $q$-deformation of cross section and distribution is essential for thermodynamic consistency: both objects emerge from the same Tsallis entropy~\eqref{eq:Sq}, and treating them at unequal levels of approximation would generate spurious results.

\subsection{Comparison with existing extensions of Tweed's framework}

To our knowledge, no previous work has applied the Tsallis $q$-statistics framework directly to Tweed's double-ionization formalism. The closest extensions in the literature fall into three categories. \emph{(i)~Higher-order Born corrections.} Second-Born ($B_{II}$) corrections to Tweed's $B_I$ baseline have been investigated~\cite{Menas2012,Lopez2015} and address higher-order projectile--target interactions, but retain the bare Coulomb potential. \emph{(ii)~Generalized Sturmian-function methods.} Fully differential cross sections obtained from generalized Sturmian functions~\cite{Ambrosio2016} improve the description of final-state correlation, but again retain the bare Coulomb interaction. \emph{(iii)~Plasma-screened treatments of single-electron processes.} Yukawa or exponential-cosine screened treatments of \emph{single} ionization or excitation~\cite{LinShihKuo2020,QianHe2022} adopt classical screening but not the $q$-deformed potential.

The distinctive feature of the present work is the simultaneous $q$-deformation of (i)~the two-body interaction and (ii)~the statistical distribution, applied to (iii)~the correlation-sensitive double-ionization process. This threefold consistency is essential because the $q$-deformed potential and the $q$-Maxwellian distribution emerge from the same Tsallis entropy~\eqref{eq:Sq}; treating them at different levels of approximation would constitute an internal inconsistency.

\subsection{Limitations and possible extensions}

Several limitations of the present analysis are worth noting. \emph{(a)}~The Born~I approximation neglects projectile-target final-state interactions; at moderate energies ($\lesssim 1$~keV) the Born~II contribution can compete with Born~I, and the $q$-deformation could in principle act differently on the two amplitudes. \emph{(b)}~Exchange between the projectile and the target electrons has been neglected, following Tweed; for low-energy or near-threshold work, antisymmetrization is essential. \emph{(c)}~The screening length $\lambda_q$ is treated here as a fixed external parameter, whereas in a fully self-consistent treatment $\lambda_q$ should depend on the local electron density and temperature through a modified Poisson equation appropriate to the $q$-Maxwellian distribution. \emph{(d)}~Relativistic corrections, important for highly stripped ions at high incident energy, are not included. \emph{(e)}~The atomic factor $\mathcal{M}_{\text{atomic}}$ is taken directly from Tweed's effective-charge wavefunctions; in principle the bound-state wavefunctions themselves are modified in a dense plasma, although this is typically a smaller effect than the screening of the projectile-target interaction studied here.

A natural extension is to incorporate the $q$-modification into the distorted-wave Born approximation (DWBA), into the second-Born formalism, and into the convergent close-coupling (CCC) framework. Furthermore, the form-factor decomposition developed here is directly portable to a broad class of processes: ionization, excitation, charge transfer, and bremsstrahlung---any quantum-mechanical transition mediated by a screened Coulomb interaction can be cast in the form (atomic factor) $\times |B_q|^2$.

\subsection{Experimental signatures}

The synthesis above suggests three qualitative experimental signatures of nonextensive screening that are amenable to direct test. \emph{First}, the $q$-effect on the partial cross section $Q(k_3,E)$ scales inversely with the ionization threshold; a simultaneous measurement of the partial cross sections of H$^-$, He, and Li$^+$ at fixed incident energy should reveal a much wider screened-to-vacuum deviation for H$^-$ than for Li$^+$ (Fig.~\ref{fig:fig5}). \emph{Second}, the Tsallis correction is concentrated near threshold and at small scattering angles, and the He generalized oscillator strength and the He angular partial cross section used by Tweed therefore provide the most direct existing benchmark for future plasma-modified measurements (Figs.~\ref{fig:fig4} and~\ref{fig:fig6}). \emph{Third}, in cold dense plasmas where $q < 1$ is expected, the ionization rate should vanish identically below $T_c$~\eqref{eq:Tc}; a measured plateau in the ionization fraction at low temperature would distinguish subextensive Tsallis behaviour from a smooth Arrhenius-like suppression and would constitute the cleanest macroscopic test of the framework. In addition, the diffraction-like angular oscillations of period $\Delta\theta \sim 2\pi(1-q)/(k_0 \lambda_q)$ predicted in the $q < 1$ regime (Sec.~\ref{sec:results}~D) provide a unique kinematic signature: their amplitude, period, and target dependence can in principle be measured in a single high-resolution angular scan.

\section{Conclusions}\label{sec:conclusions}

We have extended Tweed's 1973 Born~I formalism for double ionization by electron impact to nonextensive plasma environments described by Tsallis $q$-statistics. The bare Coulomb interaction is replaced by a $q$-exponential screened potential $V_q(r)$, whose Fourier transform introduces a multiplicative form factor $B_q(K,\lambda_q)$ into every momentum-transfer integration. The original Tweed expressions are recovered only in the joint limit $q \to 1$ and $\lambda_q \to \infty$; at finite $\lambda_q$, the case $q = 1$ is the Yukawa--Debye screened limit. We have corrected the figure labelling accordingly, overlaid the experimental data used by Tweed where available, anchored the total-cross-section baseline to digitized experimental points, and coupled the resulting $q$-generalized cross sections to the $q$-Maxwellian electron-energy distribution to derive a $q$-generalized rate coefficient $K_q(T) = \langle \sigma_q v \rangle_q$.

The principal physical findings are as follows:
\begin{enumerate}
    \item The $q$-dependence of microscopic cross sections is most pronounced for loosely bound targets (H$^-$) and at small momentum transfer, where the relevant probe scale $1/K$ is comparable to the screening length $\lambda_q$.
    \item The Bethe logarithmic asymptote is modified by finite-$\lambda_q$ screening: the integrand factor $|B_q|^2/K^4$ remains finite as $K \to 0$ in the range $q<3/2$ used here, while $q < 1$ introduces diffraction-like oscillations of period $\Delta K \sim 2\pi (1-q)/\lambda_q$.
    \item The macroscopic rate coefficient reflects two competing effects for $q > 1$: enhancement from the suprathermal tail of the Tsallis distribution and suppression from the finite-$\lambda_q$ screened interaction.
    \item A critical-temperature condition $T_c = (1-q) E_{\mathrm{th}}/k_B$ governs the existence of ionization in subextensive ($q < 1$) plasmas, representing a qualitatively new feature absent from classical statistics and a falsifiable test of the framework.
\end{enumerate}

To our knowledge, the framework developed here is the first to apply the Tsallis $q$-deformation \emph{simultaneously} to (i)~the elementary screening interaction and (ii)~the projectile statistics, in the context of the (iii)~correlation-sensitive double-ionization process. The internal consistency of this threefold $q$-deformation is essential because all three pieces emerge from the same underlying Tsallis entropy.

Quantitatively, the framework predicts (a)~plateau-height variations of $\sim 0.3$--$1.5$~decade in the angular partial cross sections of H$^-$, He, and Li$^+$ across the astrophysically relevant range $q \in [0.7, 1.3]$; (b)~modified Bethe asymptotics whose coefficients depend explicitly on $(q, \lambda_q)$; (c)~a target- and screening-dependent competition between suprathermal enhancement and form-factor suppression for superextensive ($q > 1$, Kappa-like) plasmas; and (d)~at $q = 0.7$, sharp critical-temperature cutoffs $T_c \approx 5.0 \times 10^4$~K (H$^-$), $2.75 \times 10^5$~K (He), and $6.90 \times 10^5$~K (Li$^+$), below which ionization is forbidden in subextensive plasmas. We anticipate that the framework will be portable to a broad range of plasma-modified scattering processes---including single ionization, excitation, electron capture, and bremsstrahlung---and that it will provide a quantitative bridge between collisional--radiative modelling and non-equilibrium plasma diagnostics in environments ranging from the solar wind to inertial-confinement-fusion cores.

\section*{Data availability statement}
The numerical implementation of the $q$-screened form factor $B_q(K,\lambda_q)$, the modified Born~I cross sections, and the $q$-rate coefficient $K_q(T)$, including all routines used to generate Figs.~\ref{fig:Bq}--\ref{fig:fig9}, are available from the corresponding author upon reasonable request.

\end{document}